\documentclass[aps,prb,twocolumn,superscriptaddress]{revtex4}
\usepackage{amsfonts}
\usepackage{amssymb}
\usepackage{amsmath}
\usepackage{graphicx}
\usepackage{epsfig}
\usepackage{color}
\usepackage{amsmath,bm}

\begin{document}

\title{Field induced conducting state in Mott insulator}
\author{X. Z. Zhang}
\affiliation{College of Physics and Materials Science, Tianjin Normal University, Tianjin
300387, China}
\author{Z. Song}
\email{songtc@nankai.edu.cn}
\affiliation{School of Physics, Nankai University, Tianjin 300071, China}

\begin{abstract}
Electron--electron repulsion, on the one hand, can result in bound pair,
which has heavy effective mass. On the other hand, it is also the cause of
Mott insulator. We study the effect of a staggered magnetic field on a
Hubbard model. We find that a bound pair with large energy bandwidth can be
formed under the resonant staggered field, being the half of Hubbard
repulsion strength. Accordingly, the system exhibits following dynamical
behaviors: (i) When an electric field is applied, fast bound pair Bloch
oscillation occurs, while a single electron is frozen. (ii) When a quenching
resonant field is applied to an initial antiferromagnetic Mott insulating
state, the final state becomes doublon conducting state manifested by the
non-zero $\eta$ correlator and large charge fluctuation. Our finding
indicates that the cooperation of electron-electron correlation and
modulated external field can induce novel quench dynamics.
\end{abstract}

\maketitle

\section{Introduction}

The Hubbard model represents a minimal starting point for modeling strongly
correlated systems \cite%
{Hubbard1964,Fisher1989,Georges1996,Sachdev2011,Keimer2015}. It has
relevance for a broad range of correlation phenomena, including
high-temperature superconductivity \cite{Lee2006}, metal-insulator
transition \cite{Fazekas1999}, quantum criticality \cite{Sachdev2011}, to
interacting topological states of matter \cite{Wen2007}. Over the past
decades, the Hubbard model lies at the heart of investigating the cold atom
quantum simulations since these strongly correlated systems are faithful
representations of the bosonic or fermionic Hubbard models \cite%
{Jaksch2005,Bloch2008,Bloch2012}. Recent developments in quantum simulations
of the Hubbard model suggest that these cold atom systems are particularly
suitable for studying quantum many-body dynamics \cite{Giamarchi2016}, which
is less refereed in the content of condensed matter systems. A fascinating
subject is quantum quench dynamics in the strongly correlated system. It
studies how a many-body system is prepared initially in a certain initial
state, and evolves unitarily in time following the sudden change of the
parameters to a final Hamiltonian \cite{Calabrese2006}. Dynamics induced by
a quantum quench has become an active topic of research because it poses
many fundamental questions that can also be studied by current generation
experiments.

A typical feature of quench dynamics is that the considered many-body
systems are usually far from equilibrium, which can induce many intriguing
phenomena in condensed-matter materials \cite%
{Tokura2006,Yonemitsu2008,Aoki2014,Giannetti2016}. One striking example of
recent experimental observations is the light-induced superconducting-like
properties in some high-$T_{c}$ cuprates \cite%
{Fausti2011,Hu2014,Kaiser2014,Nicoletti2014} and alkali-doped fullerides
\cite{Mitrano2016,Cantaluppi2018}, which have stimulated many theoretical
investigations \cite%
{Sentef2016,Patel2016,Knap2016,Kennes2017,Sentef2017,Babadi2017,Murakami2017,Wang2018}%
. In addition, the pump-probe experiments realized in cold-atom setups have
been also triggered an intensive theoretical investigation of
out-of-equilibrium dynamics in quantum many-body systems (see e.g. the
reviews \cite%
{Polkovnikov2011,Calabrese2016,Vasseur2016,Gogolin2016,Tomita2017}). They
have been successfully used for understanding the mechanisms of relaxation
\cite%
{Cazalilla2006,Barthel2008,Cramer2008,Calabrese2011,Mossel2012,Collura2013,Fagotti2013,Kormos2014,Piroli2017}%
, showing that local-in-space observables relax at large times. Such novel
dynamic phenomena are ultimately attributed to the cooperation between the
interaction and the external field.

This paper aims to study the effect of the staggered external field on a
Fermi-Hubbard model in the deep Mott insulating phase. We show that the
quench resonant field can build a bridge for the transport of the particles
demonstrated by the large charge fluctuation and enhancement of the
doublon-doublon correlation. Specifically, an initial anti-ferromagnetic
insulating state becomes a conducting state under the action of the
post-quench Hamiltonian. The underlying mechanism, which can be elucidated
by the two-particle system, is that: The combination between the on-site
interaction and the presence of resonant staggered field culminates in the
formation of a bound pair (BP) with large energy bandwidth. However, the
bandwidth of a single particle is shrunk. Correspondingly, the system
exhibit two distinct dynamic behaviors when the dc electric field is
applied. For the scattering states, the particles are frozen in the initial
place. As for the BPs, they show breathing behavior populating a large
interval that is determined by the bandwidth of BP. This is in stark
difference with the case of the staggered field free, wherein the BP is
frozen at the starting point. Hence, the large-scale migration of BPs
induced by the external field provides an insight into the dynamics
transition of the system from the Mott insulating to conducting phase. It is
hoped that these results can motivate further studies of both fundamental
aspects and potential applications of quench interacting many-body systems.
\begin{figure}[tbp]
\centering
\includegraphics[width=0.4\textwidth]{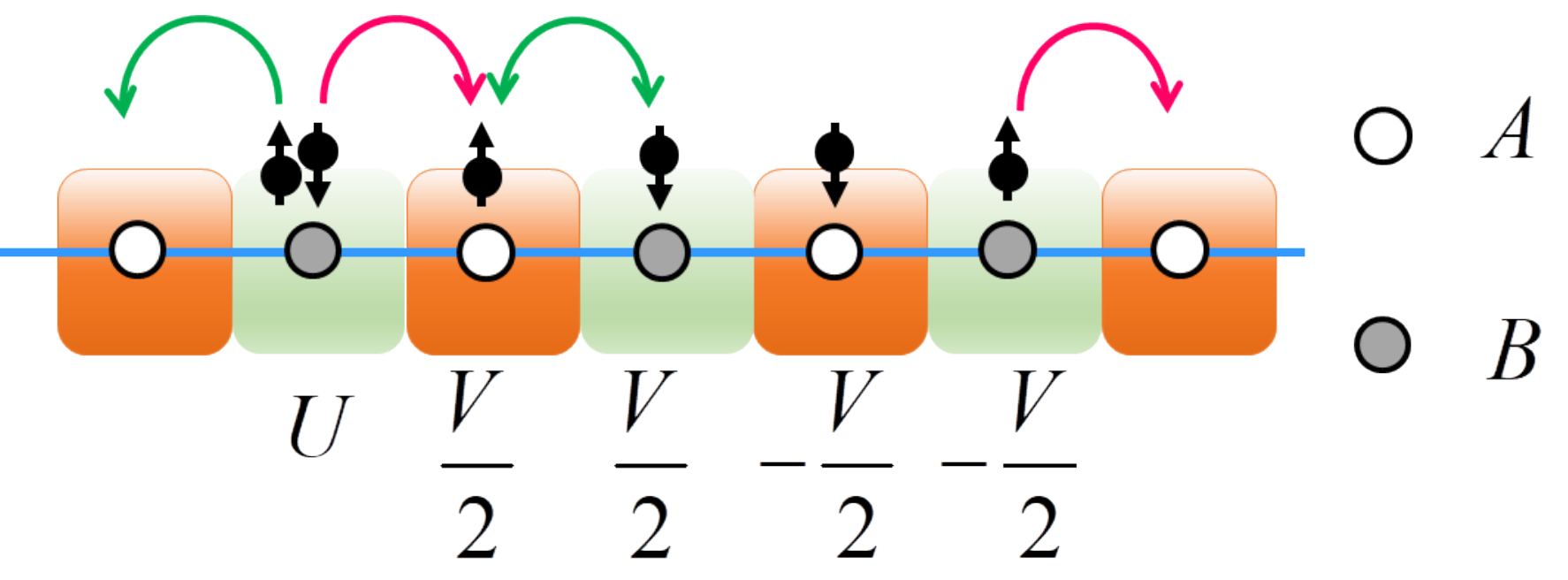}
\caption{Schematic illustration of the structure of the Hubbard model
concerned in this work, which supports the transition from insulating to
conducting states induced by the competition between on-site repulsion and
staggered magnetic field. It consists of two sub-lattices $A$ (empty) and $B$
(grey). The electrons are subjected by opposite magnetic fields which are
denoted by orange and green shaded regions, respectively. We present a
possible electron filling configuration to illustrate the underlying
mechanism of the transition. The corresponding on-site energy is denoted
under each sits. (i) In the case of $V=0$ and large $U$, the hopping
processes denoted by red arrows are allowed, while green is forbidden. The
half-filled ground state is Mott insulating state. (ii) When the resonant
magnetic field is switched on, the hopping processes denoted by green arrows
are allowed, while red is forbidden, supporting conducting state even at
half filling. }
\label{fig_illustration}
\end{figure}
The rest of the paper is structured as follows: In Sec. \ref{Model}, we
introduce the Hubbard model subjected to a staggered field. The exact
two-particle equivalent Hamiltonian and effective Hamiltonian about the BP
solution are investigated based on the symmetry of the system. Sec. \ref{BP}
is devoted to investigate the dynamics demonstrated by the BO. Physical
insight into the many-body system with a resonant staggered field is
provided by the so-called breathing oscillation of BP. In Sec. \ref{quench},
the transition from the Mott insulating to doublon conducting phase is
investigated through quench dynamics by examining the charge fluctuation and
$\eta $ correlator. We conclude and discuss our results in Sec. \ref{summary}%
.

\section{Model}

\label{Model}

\subsection{the Hubbard model with staggered potential}

We consider a many-body quantum system $H=H_{0}+H_{s}$ and show the basic
property determined by the interplay between the interaction and external
field. The system Hamiltonian given by the Hubbard model with bipartite
lattice
\begin{equation}
H_{0}=-\sum_{i\in A,j\in B}^{N}\sum_{\sigma =\uparrow ,\downarrow
}J_{ij}c_{i,\sigma }^{\dagger }c_{j,\sigma }+\text{H.c.}+U\sum_{j\in A\cup
B}n_{j,\uparrow }n_{j,\downarrow },  \label{H0}
\end{equation}%
is subjected to a staggered external field
\begin{equation}
H_{s}=\frac{V}{2}\sum_{j\in A}\left( n_{j,\uparrow }-n_{j,\downarrow
}\right) -\frac{V}{2}\sum_{j\in B}\left( n_{j,\uparrow }-n_{j,\downarrow
}\right) .
\end{equation}%
where the operator $c_{j,\sigma }$ is the annihilation operator of a spin-$%
\sigma $ fermion at site $j$, satisfying the usual fermion anticommutation
relations $\{c_{i,\sigma }^{\dagger },$ $c_{j,\sigma ^{\prime }}\}=\delta
_{i,j}\delta _{\sigma ,\sigma ^{\prime }}$ and $\{c_{i,\sigma },$ $%
c_{j,\sigma ^{\prime }}\}=0$; the system can be divided into two disjoint
sets $A$ and $B$ and the hopping matrix elements $J_{ij}$ are required to be
real and satisfy $J_{ij}=J_{ji}$; the number operators are $n_{j,\sigma
}=c_{j,\sigma }^{\dagger }c_{j,\sigma }$, while $U$ is the on-site energy
denoting the repulsive particle-particle interaction; the term $V$ describes
spin-dependent staggered potentials. For convenience and clarity, the number
of the sublattice sites and the filled-particles are denoted by $N$ and $M$,
respectively. The considered system is illustrated in Fig. \ref%
{fig_illustration}. Here, we want to stress that the staggered magnetic
field can alter significantly the mobility of the particles. A simple
example is employed to elucidate the underlying mechanism in Fig. \ref%
{fig_illustration. It is shown that the external staggered field provides
an energy shell that allows particle pairs to move through more channels,
which may facilitate the enhancement of the conductivity. In the subsequent
sections, we will gradually dissect the influence of external magnetic field
on the energy spectrum and corresponding dynamics. To this end, we first
focus on the symmetry of the system.} In the absence of the staggered field $%
V=0$, the system Hamiltonian $H$ has two sets of SU(2) symmetry. The first
of these relates to spinful particles which can be characterized by the
generators
\begin{eqnarray}
s^{+} &=&\left( s^{-}\right) ^{\dagger }=\sum_{j\in A\cup B}s_{j}^{+}, \\
s^{z} &=&\sum_{j\in A\cup B}s_{j}^{z},
\end{eqnarray}%
where the local operators $s_{j}^{+}=c_{j,\uparrow }^{\dagger
}c_{j,\downarrow }$ and $s_{j}^{z}=\left( n_{j,\uparrow }-n_{j,\downarrow
}\right) /2$ obey the Lie algebra, i.e., $[s_{j}^{+},$ $%
s_{j}^{-}]=2s_{j}^{z} $, and $[s_{j}^{z},$ $s_{j}^{\pm }]=\pm s_{j}^{\pm }$.
The second, often referred to as $\eta $ symmetry, relates to spinless
quasiparticles. The corresponding generators can be given as
\begin{eqnarray}
\eta ^{+} &=&\left( \eta ^{-}\right) ^{\dagger }=\sum_{j\in A\cup B}\eta
_{j}^{+}, \\
\eta ^{z} &=&\sum_{j\in A\cup B}\eta _{j}^{z},
\end{eqnarray}%
with $\eta _{j}^{+}=\lambda c_{j,\uparrow }^{\dagger }c_{j,\downarrow
}^{\dagger }$ and $\eta _{j}^{z}=\left( n_{j,\uparrow }+n_{j,\downarrow
}-1\right) /2$ satisfying commutation relation, i.e., $[\eta _{j}^{+},$ $%
\eta _{j}^{-}]=2\eta _{j}^{z}$, and $[\eta _{j}^{z},$ $\eta _{j}^{\pm }]=\pm
\eta _{j}^{\pm }$. Here we assume a bipartite lattice and $\lambda =1$ for $%
j\in \left\{ A\right\} $ and $-1$ for $j\in \left\{ B\right\} $. Evidently,
the presence of the staggered field spoils such SU(2) symmetries but remains
$s^{z}$ symmetry, that is $[s^{z},$ $H]=0$. Hence, the system can be divided
into different invariant subspaces in terms of the quantum number of $s^{z}$.

\subsection{bound pair solution}

\begin{figure}[tbp]
\centering
\includegraphics[width=0.5\textwidth]{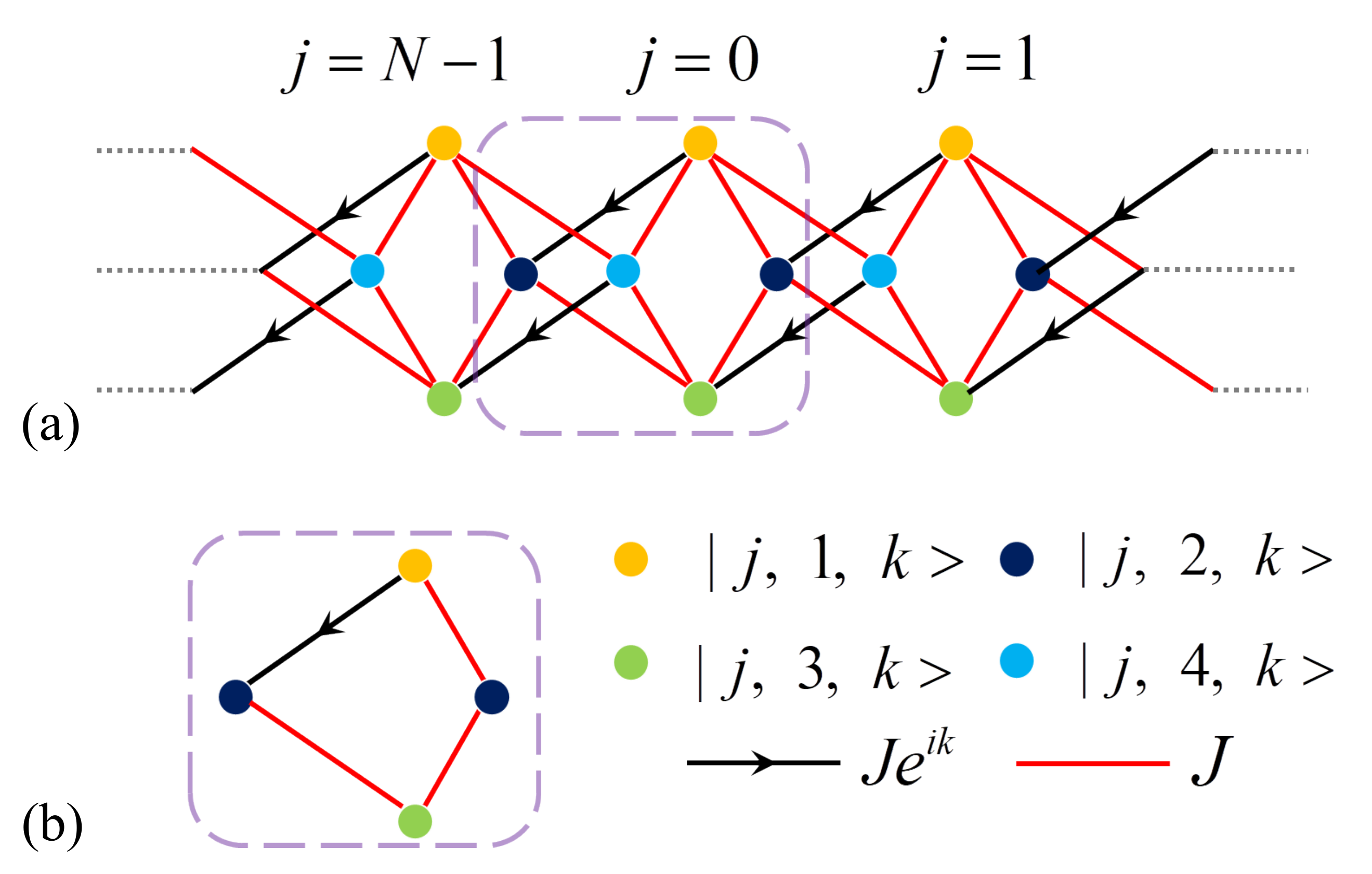}
\caption{Schematic of the structures of equivalent Hamiltonians for the
one-dimensional two-particle Hubbard model with the staggered field. (a) In
the invariant subspace with center momentum $k$, the equivalent Hamiltonian $%
H_{k}$ describes a $4$-leg ladder with $k$-dependent hopping. The on-site
interaction can be treated as the on-site potential that only exists in the $%
j=0$ unit cell. The staggered field appears periodically within each unit
cell. (b) The purple box describes the effective Hamiltonian about the BP
approximate solution. In the near-resonant condition and large $U$ limit, $%
H_{\mathrm{eff}}$ characterizes the first-order perturbation.}
\label{fig1}
\end{figure}
\begin{figure*}[tbp]
\centering
\includegraphics[width=0.9\textwidth]{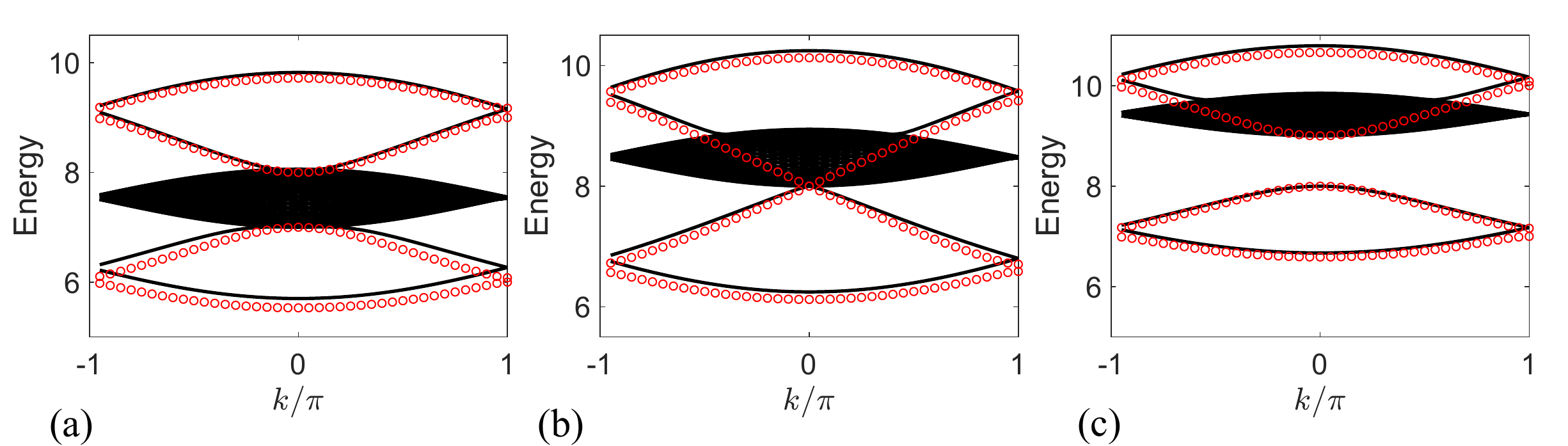}
\caption{Comparison between the approximate BP solution and exact
diagonalization results. There are three typical cases: (a) $U/J=8$, $V/J=7$%
; (b) $U/J=8$, $V/J=8$; (c) $U/J=8$, $V/J=9$; One can see that when a large $%
U$ limit takes, two such results accord well with each other. $H_{eff}$ can
well capture the information about the touching between the upper and lower
BP band. The black shaded region represents the scattering band, which can
be obtained by the single-particle solution.}
\label{fig2}
\end{figure*}

To gain physical insight into subsequent many-body dynamics, we first
investigate the two-particle invariant subspaces. When $V=0$, the 1D
homogeneous system can harbor the bound pair (BP) solution with the form of $%
\varepsilon _{\mathrm{bp}}^{k}=\sqrt{U^{2}+16J^{2}\cos ^{2}\frac{k}{2}}$
\cite{Winkler2006,Jin2011,Zhang2016,Zhang2021a}. The corresponding bandwidth
is $8J^{2}/U$ as $J\ll U$. This becomes very complicated when staggered
field presents. For clarity and simplicity, we suppose that the Hamiltonian $%
H$ still describes a 1D homogeneous ring system. The translation symmetry
combing with $s^{z}$ symmetry allows us to construct the basis of invariant
subspace $s^{z}=0$ as follow%
\begin{equation}
\left(
\begin{array}{c}
\left\vert j,1,k\right\rangle \\
\left\vert j,2,k\right\rangle \\
\left\vert j,3,k\right\rangle \\
\left\vert j,4,k\right\rangle%
\end{array}%
\right) =\sum_{l}\frac{1}{\sqrt{N}}e^{ik(l+j)}\left(
\begin{array}{c}
\alpha _{l,\uparrow }^{\dagger }\alpha _{l+j,\downarrow }^{\dagger }|\mathrm{%
Vac}\rangle \\
\alpha _{l,\uparrow }^{\dagger }\beta _{l+j,\downarrow }^{\dagger }|\mathrm{%
Vac}\rangle \\
\beta _{l,\uparrow }^{\dagger }\beta _{l+j,\downarrow }^{\dagger }|\mathrm{%
Vac}\rangle \\
\beta _{l,\uparrow }^{\dagger }\alpha _{l+j,\downarrow }^{\dagger }|\mathrm{%
Vac}\rangle%
\end{array}%
\right) ,
\end{equation}%
with the notation $\alpha _{l,\sigma }^{\dagger }=c_{l,\sigma }^{\dagger }$ (%
$l\in \left\{ A\right\} $) and $\beta _{l,\sigma }^{\dagger }=c_{l,\sigma
}^{\dagger }$ ($l\in \left\{ B\right\} $). Here $k=2n\pi /N$ is the momentum
vector indexing the subspace and $j=0...N-1$ describes the relative distance
between the two particles. The other two invariant subspaces $s^{z}=\pm 1$
are not considered since the bound states only reside in the subspace with $%
s^{z}=0$. In such a two-particle Hilbert space, the Hamiltonian $H$ can be
written as $H=\sum_{k}H_{k}$ where
\begin{eqnarray}
H_{k} &=&\sum_{j=0}^{N-1}h_{j,k}-J\sum_{j=0}^{N-1}\sum_{m}\left\vert
j,m,k\right\rangle \left\langle j,m+1,k\right\vert +\text{H.c.}  \notag \\
&&+\sum_{j=0}^{N-1}\sum_{m}\mu _{m}\left\vert j,m,k\right\rangle
\left\langle j,m,k\right\vert +  \notag \\
&&+U(\left\vert 0,1,k\right\rangle \left\langle 0,1,k\right\vert +\left\vert
0,3,k\right\rangle \left\langle 0,3,k\right\vert ),  \label{Hk}
\end{eqnarray}%
and%
\begin{eqnarray}
h_{j,k} &=&\lambda _{1}\left\vert j,1,k\right\rangle \left\langle
j+1,4,k\right\vert +\lambda _{2}\left\vert j,2,k\right\rangle \left\langle
j+1,1,k\right\vert  \notag \\
&&+\lambda _{3}\left\vert j,2,k\right\rangle \left\langle j+1,3,k\right\vert
\notag \\
&&+\lambda _{4}\left\vert j,3,k\right\rangle \left\langle j+1,4,k\right\vert
+\text{H.c.}
\end{eqnarray}%
with $\lambda _{1,2,3,4}=-J\left( 1,\text{ }e^{ik},\text{ }1,\text{ }%
e^{ik}\right) $ and $\mu _{1,2,3,4}=\left( 0,\text{ }V,\text{ }0,\text{ }%
-V\right) $. Note that $\left\vert N,m,k\right\rangle =\left\vert
0,m,k\right\rangle $. The expression of $H_{k}$ in Eq. (\ref{Hk}) has a
clear physical meaning: $j$, $m$, $k$ denotes the site state for the $j$-th
site on the $m$-leg of a $4$-leg ladder system with the modulated inter-cell
coupling. The structure of $H_{k}$ is schematically illustrated in Fig. \ref%
{fig1}(a). The BP state mainly piles up at the $0$-th site due to the near
resonance condition of $\left\vert \Delta \right\vert /J\ll 1$, where $%
\Delta =U-V$. Hence, the corresponding approximate solution can be obtained
within such unitcell. The involved four kets are denoted by purple box in
Fig. \ref{fig1}(b). In the basis of $\left\{ \left\vert 0,1,k\right\rangle
\text{, }\left\vert 0,2,k\right\rangle \text{, }\left\vert
0,3,k\right\rangle \text{, }\left\vert N-1,2,k\right\rangle \right\} $, the
matrix form of effective BP Hamiltonian is given as
\begin{equation}
H_{\mathrm{eff}}=-J\left(
\begin{array}{cccc}
-\Delta /J & 1 & 0 & e^{-ik} \\
1 & 0 & 1 & 0 \\
0 & 1 & -\Delta /J & 1 \\
e^{ik} & 0 & 1 & 0%
\end{array}%
\right) +VI_{4},
\end{equation}%
Taking the linear transformation,
\begin{equation}
\left(
\begin{array}{c}
\left\vert 1,k\right\rangle _{\mathrm{bp}} \\
\left\vert 2,k\right\rangle _{\mathrm{bp}} \\
\left\vert 3,k\right\rangle _{\mathrm{bp}} \\
\left\vert 4,k\right\rangle _{\mathrm{bp}}%
\end{array}%
\right) =\frac{1}{\sqrt{2}}\left(
\begin{array}{cccc}
-1 & 0 & 1 & 0 \\
0 & -1 & 0 & 1 \\
1 & 0 & 1 & 0 \\
0 & 1 & 0 & 1%
\end{array}%
\right) \left(
\begin{array}{c}
\left\vert 0,1,k\right\rangle \\
\left\vert 0,2,k\right\rangle \\
\left\vert 0,3,k\right\rangle \\
\left\vert N-1,2,k\right\rangle%
\end{array}%
\right) ,
\end{equation}%
the effective BP Hamiltonian can be rewritten as%
\begin{equation}
H_{\mathrm{eff}}=\left(
\begin{array}{cc}
h_{1,\mathrm{eff}} & 0 \\
0 & h_{2,\mathrm{eff}}%
\end{array}%
\right) +VI_{4},
\end{equation}%
with
\begin{eqnarray}
h_{1,\mathrm{eff}} &=&-J\left(
\begin{array}{cc}
\Delta /J & 2i\sin \left( k/4\right) \\
-2i\sin \left( k/4\right) & 0%
\end{array}%
\right) , \\
h_{2,\mathrm{eff}} &=&-J\left(
\begin{array}{cc}
\Delta /J & -2\cos \left( k/4\right) \\
-2\cos \left( k/4\right) & 0%
\end{array}%
\right) .
\end{eqnarray}%
Evidently, the effective $H_{eff}$ can be divided into two invariant
subspaces, each of which is described by a $k$-dependent two-level system.
The corresponding eigen energies are
\begin{eqnarray}
\varepsilon _{1,\sigma }^{k} &=&\frac{\Delta }{2}+\sigma \sqrt{(\frac{\Delta
}{2})^{2}+4J^{2}\sin ^{2}\frac{k}{4}},  \label{eff_1} \\
\varepsilon _{2,\sigma }^{k} &=&\frac{\Delta }{2}+\sigma \sqrt{(\frac{\Delta
}{2})^{2}+4J^{2}\cos ^{2}\frac{k}{4}},  \label{eff_2}
\end{eqnarray}%
with $\sigma =\pm $. For each subsystem, there exists an energy gap $\Delta $
between the two bands. When the resonance condition ($\Delta =0$) is
achieved, not only the upper and lower energy bands are connected with each
other ($\varepsilon _{j,+}^{0}=\varepsilon _{j,-}^{0}$, $j=1,$ $2$), but
also four bands form a single BP band with the width $4J$. This evidence
suggests that the bound pair velocity is altered from the order of $J^{2}/U$
to $J$. To check this point, we compare the result obtained by Eqs. (\ref%
{eff_1})-(\ref{eff_2}) with that of the exact diagonalization method in Fig. %
\ref{fig2}. It demonstrates that two kinds of results accord with each other
in the large $U$ limit. This characteristic plays the key role to under the
following Bloch oscillation (BO) dynamics.

\section{dynamics of two-particle system}

\label{BP} It is well known that a particle in a periodic potential with an
additional constant force performs BO, which can be explained by the
formation of a Wannier-Stark ladder energy spectrum. BO is an ubiquitous
phenomenon of coherent wave transport that can be observed in a wide variety
of physical system \cite%
{Voisin1988,Feldmann1992,BenDahan1996,Wilkinson1996,Morandotti1999,Pertsch1999,Rosam2001,Morsch2001,Longhi2008,Dreisow2009,Zhang2017}
and reflects the spectrum property of the system in the absence of the weak
dc electric field \cite{Hartmann2004,Breid2006}. In addition, a cluster of
bound particles can also present the BO which shares the same mechanism of
the single-particle case \cite{Lin2014,Zhang2016}. In this section, we use
the BO to investigate the influence of a staggered field on the transport
properties of the system.
\begin{figure*}[tbp]
\centering
\includegraphics[width=0.9\textwidth]{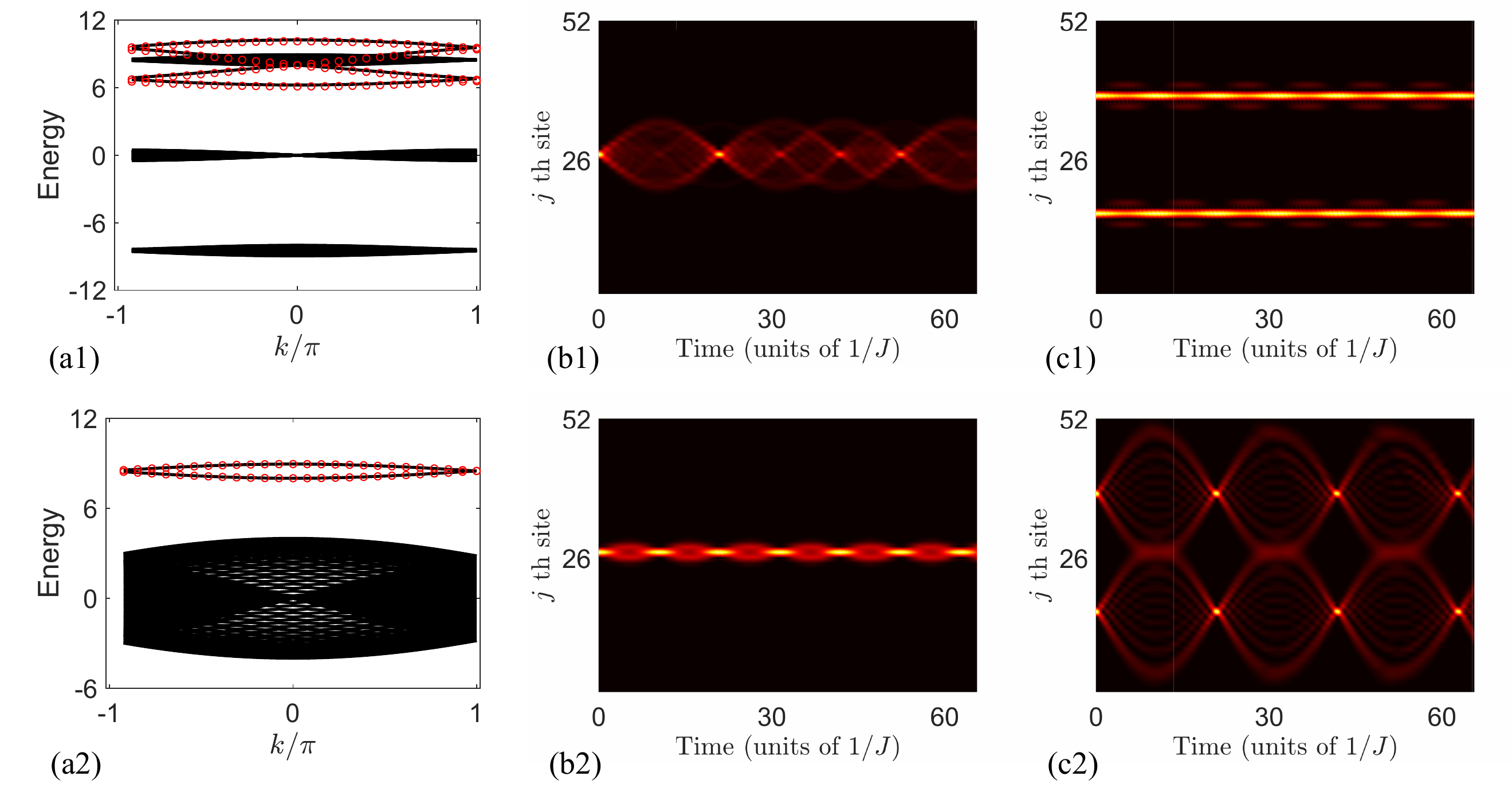}
\caption{Band structures for the systems with and without staggered field $V$%
: (a1) $U/J=8$, $V/J=7$; (a2) $U/J=8$, $V=0$. The presence of $V$ results in
the two-band structure for the single-particle subspace so that the
scattering bands of (a1) consists of three parts with different energy
scale. Such the scattering band can be understood by the superposition of
the energies of two-free particles. The resonant staggered field expands the
BP band but suppresses the scattering band. The different band structures
determine the system dynamics in the presence of a dc electric field. The
panels (b1)-(b2) describes the BP dynamics, while the (c1)-(c2) exhibit the
time evolution of the two scattering-particle state $\left\vert \Phi
_{1}\left( 0\right) \right\rangle $. It clearly shows that the scattering
particles are frozen at their initial locations and the BP can widen and
shrink periodically populating an interval $4J/F$ when dc electric field is
applied. This indicates that the BP band induced by staggered field is $4J$,
which again confirms our analysis.}
\label{fig3}
\end{figure*}

\subsection{single-electron solution: Rice-Mele model}

We first study the 1D bipartite Hamiltonian in the single-particle invariant
subspace, wherein the interaction $U$ has no effect. On the basis of
\begin{eqnarray}
\left\vert \lambda ,j,+\right\rangle &=&c_{j,\uparrow }^{\dagger }\left\vert
\mathrm{Vac}\right\rangle , \\
\left\vert \lambda ,j,-\right\rangle &=&c_{j,\downarrow }^{\dagger
}\left\vert \mathrm{Vac}\right\rangle ,
\end{eqnarray}%
where $\left\vert \mathrm{Vac}\right\rangle $ is the vacuum state of fermion
$c_{j,\sigma }$, and $\lambda =A$ ($B$) when $j\in A$ ($B$), the
corresponding Hamiltonian can be given as%
\begin{equation}
H=H_{0}+H_{s}+H_{e},
\end{equation}%
with%
\begin{eqnarray}
H_{0} &=&-J\sum_{j,\sigma }\left( \left\vert A,j,\sigma \right\rangle
\left\langle B,j,\sigma \right\vert \right.  \notag \\
&&\left. +\left\vert B,j,\sigma \right\rangle \left\langle A,j+1,\sigma
\right\vert +\text{H.c.}\right) , \\
H_{s} &=&-\frac{V}{2}\sum_{j,,\sigma }\left( \sigma \left\vert A,j,\sigma
\right\rangle \left\langle A,j,\sigma \right\vert \right.  \notag \\
&&\left. -\left\vert B,j,\sigma \right\rangle \left\langle B,j,\sigma
\right\vert \right) , \\
H_{e} &=&F\sum_{j,\sigma }\left( 2j+1\right) \left\vert A,j,\sigma
\right\rangle \left\langle A,j,\sigma \right\vert  \notag \\
&&+2j\left\vert B,j,\sigma \right\rangle \left\langle B,j,\sigma \right\vert
.
\end{eqnarray}%
Here $H_{e}$ describes the dc electric field. In the absence of $H_{d}$, the
Hamiltonian characterizes a Rice-Mele model with homogeneous hopping \cite%
{Rice1982}. Performing the Fourier transformation,
\begin{equation}
\left\vert \lambda ,q,\sigma \right\rangle =\frac{1}{\sqrt{N}}%
\sum_{j}e^{iqj}\left\vert \lambda ,j,\sigma \right\rangle ,
\end{equation}%
where $q=2l\pi /N$, $l\in \left[ 0,\text{ }N-1\right] $, the Hamiltonian $H$
can be block diagonalized due to its translational symmetry, i.e., $%
H=\sum_{q,\sigma }H_{q,\sigma }$. In this new basis, the matrix form of $%
H_{q}$ is
\begin{equation}
H_{q,\sigma }=-J\left(
\begin{array}{cc}
\sigma V/2J & 1+e^{ik} \\
1+e^{-ik} & -\sigma V/2J%
\end{array}%
\right) ,
\end{equation}%
which admits the eigen spectrum as
\begin{equation}
\epsilon _{q,\pm }^{\left( 1\right) }=\pm \sqrt{(\frac{V}{2})^{2}+4J^{2}\cos
^{2}\left( q/2\right) }.  \label{single_particle}
\end{equation}%
Evidently, the two-particle spectrum consists of scattering and bound bands.
The scattering band can be theoretically constructed through the Eq. (\ref%
{single_particle}). In particular, the upper scattering band of Figs. \ref%
{fig3}(a1)-(a2) can be obtained by the combination of two individual
particles, i.e., $\epsilon _{k,\text{\textrm{u}}}^{\left( 2\right)
}=\epsilon _{q_{1},+}^{\left( 1\right) }+\epsilon _{q_{2},+}^{\left(
1\right) }$. Analogously, the middle and lower scattering bands can be given
as $\epsilon _{k,\text{\textrm{m}}}^{\left( 2\right) }=\epsilon
_{q_{1},+}^{\left( 1\right) }+\epsilon _{q_{2},-}^{\left( 1\right) }$, and $%
\epsilon _{k,\text{\textrm{l}}}^{\left( 2\right) }=\epsilon
_{q_{1},-}^{\left( 1\right) }+\epsilon _{q_{2},-}^{\left( 1\right) }$,
respectively. When $J\ll V$, the single-particle bandwidth is in proportion
to $J^{2}/U$ which is similar to that of the bound pair spectrum as $V=0$.
This indicates that the presence of the staggered field suppresses the group
velocity of the single particle but enhances the bound pair group velocity.
It is conceivable that when $V$ is large enough, the existence of dc
electric field will not induce the large-scale oscillation of scattering
particles in coordinate space. In the following, we examine the dynamics of
the system in the single-particle and two-particle Hilbert spaces,
respectively.

\subsection{Bloch oscillation}

In the preceding section, the effect of the staggered field on the band
structure is discussed. For simplicity, we only consider the dynamics when
the resonant field is applied. The Hamiltonian can be given as $H_{\mathrm{BO%
}}=H+H_{e}$, where the dc electric field is $H_{e}=F\sum_{\sigma ,j\in
A,B}jn_{j,\sigma }$. To show the difference between the dynamics of the
single-particle and two-particle cases, the initial states are chosen as%
\begin{eqnarray}
\left\vert \Phi _{1}\left( 0\right) \right\rangle &=&\frac{1}{\sqrt{2}}%
c_{N/2,\uparrow }^{\dagger }c_{3N/2,\downarrow }^{\dagger }\left\vert
\mathrm{Vac}\right\rangle , \\
\left\vert \Phi _{2}\left( 0\right) \right\rangle &=&\frac{1}{\sqrt{2}}%
c_{N,\uparrow }^{\dagger }c_{N,\downarrow }^{\dagger }\left\vert \mathrm{Vac}%
\right\rangle .
\end{eqnarray}%
These are two types of wave packets, which are initially located at two
separate sites and a single site, respectively. Evidently, the existence of
interaction $U$ only affects $\left\vert \Phi _{2}\left( 0\right)
\right\rangle $ but does no effect on $\left\vert \Phi _{1}\left( 0\right)
\right\rangle $. For the initial state $\left\vert \Phi _{1}\left( 0\right)
\right\rangle $, the two-particle dynamics can be understood as the
superposition of the motions of two independent particles regardless of
whether the staggered external field is switched on. For the second type of
initial state $\left\vert \Phi _{2}\left( 0\right) \right\rangle $, the
interaction $U$ binds the two particles together so that the two particles
behave like a single composite particle. The dc electric field forces the
spectrum of the system to consists of Wannier--Stark ladders \cite%
{Hartmann2004,Breid2006} and hence drives the two initial states to undergo
the BO, respectively. A typical breathing behavior can be observed in Figs. %
\ref{fig3}(b1) and (c2). It shows that the width of the wave packet
oscillates strongly, while its position remains constant. In general, the
dynamics is confined to a spatial interval, the width of which is determined
by the ratio of the bandwidth of the considered system with $H_{e}=0$ to the
strength $F$. Specifically, the BP\ can be moved on the lattice with
effective hopping $2J^{2}/U$ when $V=0$ and large $U$ limit are supposed,
whereas a single particle's hopping is $J$. As such, the breathing mode of
two scattering particles $\left\vert \Phi _{1}\left( t\right) \right\rangle $
extends over an interval $4J/F$, and $\left\vert \Phi _{2}\left( t\right)
\right\rangle $ is frozen at the initial site. Note that the period of BO
for BP is $\pi /2F$ rather than $\pi /F$ since the external field felt by
the BP is $2F$ when it migrates. In contrast, the existence of the resonant
field shrinks the bandwidth of the scattering band to the order of $J^{2}/U$%
. On the other hand, it broadens the width of the bound band to $4J$ instead
of $J^{2}/U$ with frequency $F$. As a consequence, the dynamics of the BP
and scattering particles are interchanged as shown in Figs. \ref{fig3}
(c1)-(c2). So far, we have demonstrated how the external resonance field can
improve the mobility of BP in the matter by changing the structure of the
energy spectrum through the example of two particles. This mechanism could
provide some physical insights to understand the subsequent many-body
dynamics.

\section{Quench dynamics at half filling}

\label{quench} In this section, we turn to investigate the many-body quench
dynamics. Before quench, the Hamiltonian of the considered system is $H_{0}$%
, which is a standard Hubbard Hamiltonian. When $J\ll U$, the system is in
the Mott insulating phase such that there exists an excitation gap in the
spectrum. In the following, we focus on the system at half filling. In this
condition, each site contains only one particle. The on-site interaction
prohibits the movement of the particles. With the spirit of the two-particle
case, one can imagine that the resonant staggered field provides a bridge
for the mobility of the particles since the basis of \{$c_{j,\uparrow
}^{\dagger }c_{j,\downarrow }^{\dagger }\left\vert \mathrm{Vac}\right\rangle
$, $c_{j,\uparrow }^{\dagger }c_{j+1,\downarrow }^{\dagger }\left\vert
\mathrm{Vac}\right\rangle $, $c_{j+1,\uparrow }^{\dagger }c_{j+1,\downarrow
}^{\dagger }\left\vert \mathrm{Vac}\right\rangle $,...\} stays at the same
energy shell ($U$). Therefore, the conductivity of the system is enhanced.
\begin{figure}[tbp]
\centering
\includegraphics[width=0.35\textwidth]{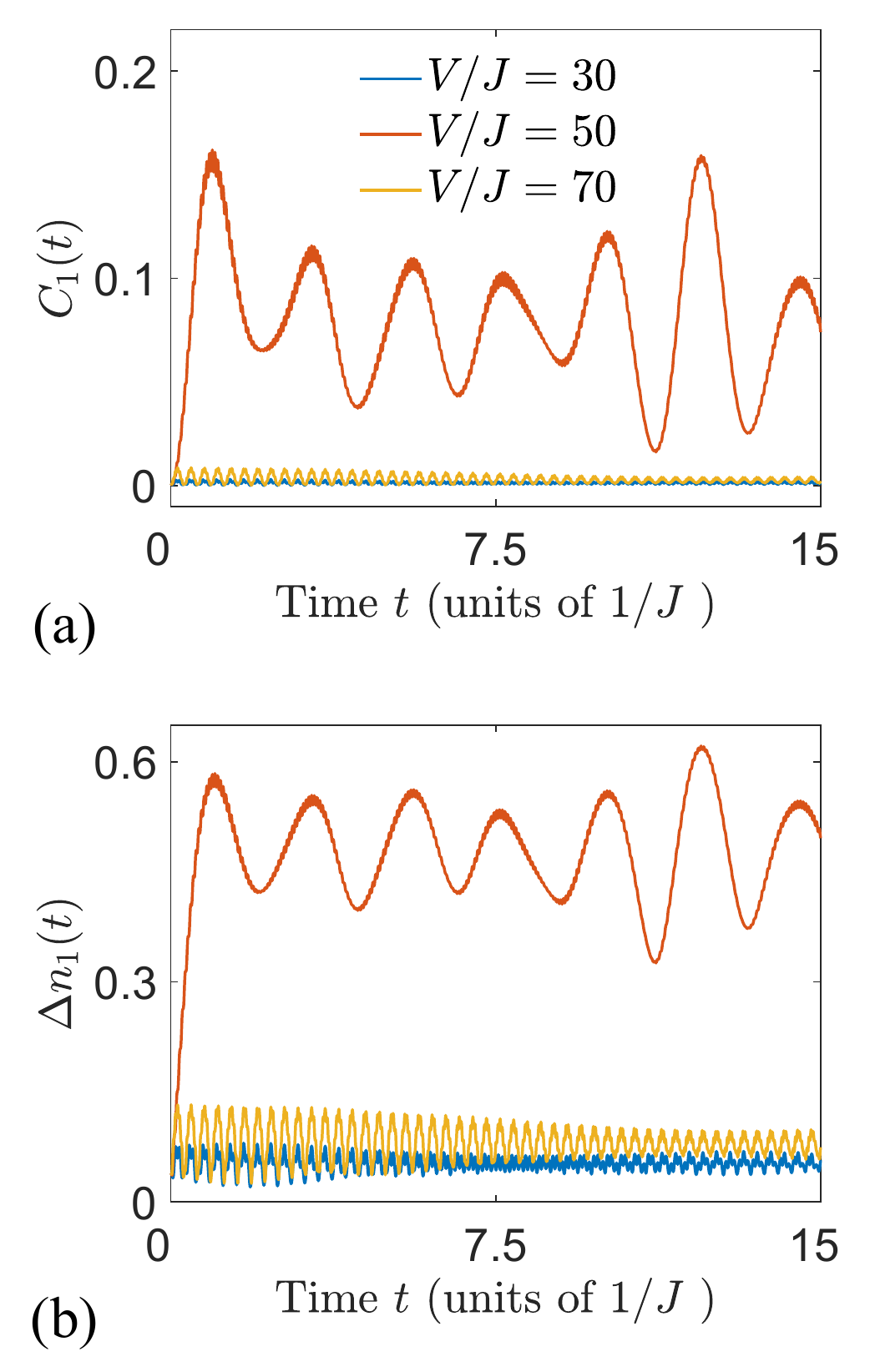}
\caption{Time evolutions of correlator $C_{r}\left( t\right) $ and charge
fluctuation $\Delta n_{j}\left( t\right) $ for different staggered fields $V$
(units of $J$) with a given on-site interaction $U/J=50$. The system is
initialized in the ground state of $H_{0}$, which is $6$-site Hubbard model
at half filling and $S_{z}=0$. Evidently, the resonant field can induce the
oscillation behaviors of both quantities. It can be served as a dynamic
signature to identify the transition from the Mott insulating to conducting
state.}
\label{fig4}
\end{figure}

\subsection{Transition from Mott insulating to conducting state}

To see how does the staggered field affect the conductivity, we propose a
quench scheme in the following. The scheme is that: The system is
initialized in the ground state of $H_{0}$ and then evolved with a
post-quench Hamiltonian $H=H_{0}+H_{s}$. To measure the conductivity of the
system, two physical quantities are introduced. The first one is the
doublon-doublon correlation
\begin{equation}
C_{i,j}=\langle \eta _{i}^{+}\eta _{j}^{-}\rangle \text{ }\forall \text{ }i%
\text{, }j\text{, }i\neq j,
\end{equation}%
the constant number of which denotes that the corresponding state has
off-diagonal long-range order (ODLRO). It also provides a possible
definition of superconductivity, as a finite value of this quantity can be
shown to imply both the Meissner effect and flux quantization \cite{Yang1962}%
. In general, the $\eta $-pairing state of $H_{0}$ possesses ODLRO.
Unfortunately, such state is the excited state rather than the ground state
of $H_{0}$. In addition, the presence of $H_{s}$ spoils the $\eta $ symmetry
so that such superconducting properties of the system may not exist.
However, the correlator $C_{i,j}$ can reflect the conductivity of the system
to some extent. The other quantity used to examine the conductivity is the
charge fluctuation $\Delta n_{j}$ defined by%
\begin{equation}
\Delta n_{j}=\sqrt{\langle (n_{j})^{2}\rangle -\langle n_{j}\rangle ^{2}}.
\end{equation}%
It is well known that when the on-site repulsive interaction between
fermions is large enough in the Mott phase, the number fluctuation would
become energetically unfavorable, forcing the system into a number state and
exhibiting a vanishing particle charge fluctuation $\Delta n_{j}$. Beyond
the Mott insulator regime, the fermions are delocalized with the
nonvanishing charge fluctuation. For the half-filled free fermion system,
straightforward algebras show that $\Delta n_{j}=\sqrt{2}/2$, which can be
served as a benchmark to examine the following result.
\begin{figure}[tbp]
\centering
\includegraphics[width=0.4\textwidth]{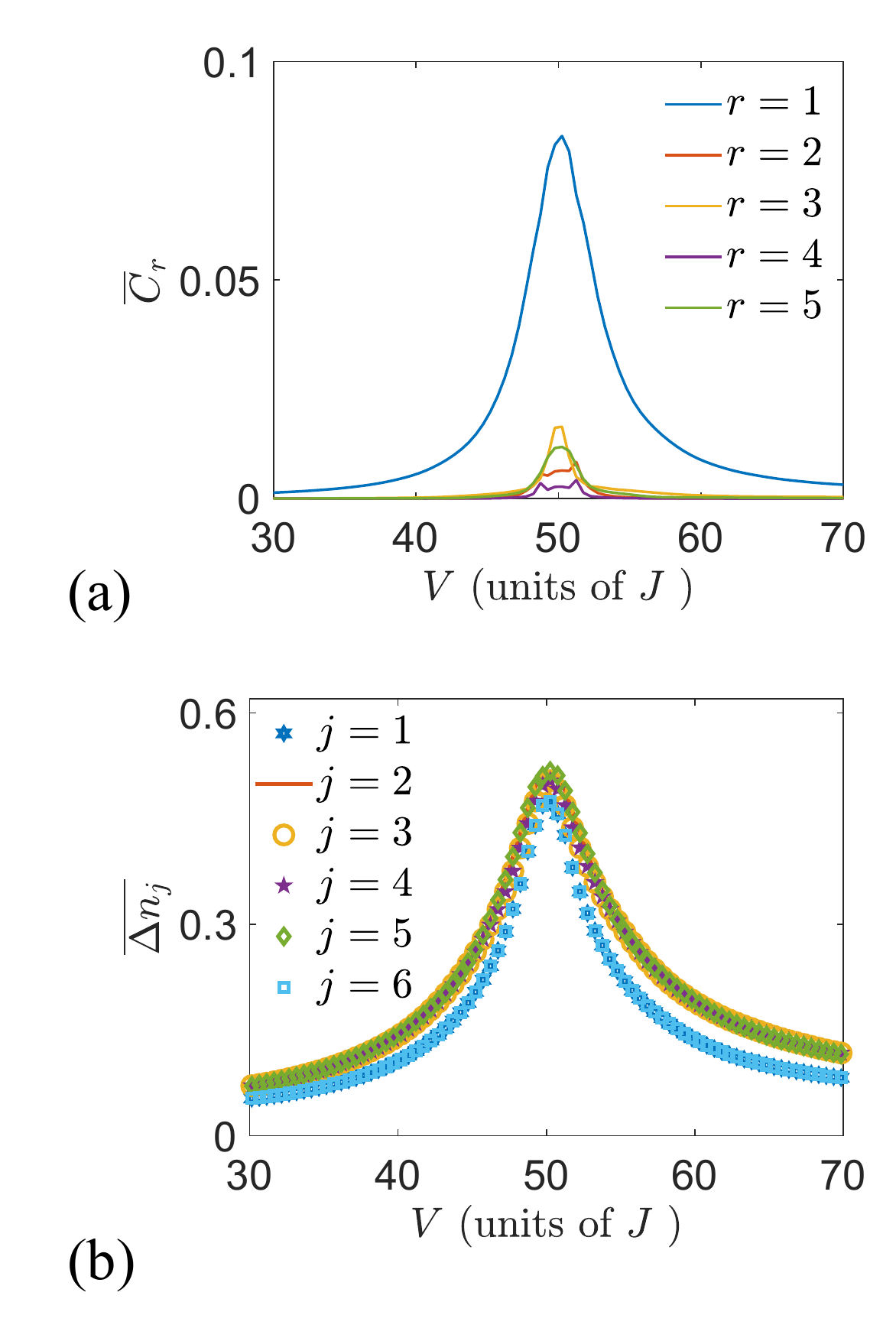}
\caption{Average $\overline{C}_{r}$ and $\overline{\Delta n_{j}}$ as
functions of $V$ for $6$-site Hubbard model at half filling and $S_{z}=0$.
The system parameters are $U/J=50$, and $t_{f}=15/J$. It indicates that as $%
V $ approaches $U$, a peak will appear demonstrating our prediction in the
text.}
\label{fig5}
\end{figure}

Again, we focus on $1$D lattice realizations due to their numerical
tractability. We perform numerical calculation where a $6$-site Hubbard
model at half filling is considered. Taking the ground state of $H_{0}$ as
the initial state $\left\vert \Phi \left( 0\right) \right\rangle $, one can
observe the variations of the correlator $C_{r}\left( t\right) =\langle \Phi
\left( t\right) |\eta _{i}^{+}\eta _{i+r}^{-}\left\vert \Phi \left( t\right)
\right\rangle $ and $\Delta n_{j}\left( t\right) =\sqrt{\langle \Phi \left(
t\right) |n_{j}^{2}\left\vert \Phi \left( t\right) \right\rangle -\left(
\langle \Phi \left( t\right) |n_{j}\left\vert \Phi \left( t\right)
\right\rangle \right) ^{2}}$ as time, where $\left\vert \Phi \left( t\right)
\right\rangle =e^{-iHt}\left\vert \Phi \left( 0\right) \right\rangle $. Note
that the ground state $\left\vert \Phi \left( 0\right) \right\rangle $ of $1$%
D Hubbard model is the Mott insulator with strong antiferromagnetic
correlations. Fig. \ref{fig4} shows that the value of $V$ which we quench to
determines the center value of two such oscillated quantities. When the
resonant field is applied, i.e. $V=U$, the maximum center values of $%
C_{r}\left( t\right) $ and $\Delta n_{j}\left( t\right) $ can be reached,
which is in agreement with our previous analysis. To determine the effect of
$V$, we introduce the average correlator and charge fluctuation in the time
interval $\left[ 0,\text{ }t_{f}\right] $, defined as follows%
\begin{eqnarray}
\overline{C}_{r} &=&\frac{1}{t_{f}}\int_{0}^{t_{f}}C_{r}\left( t\right)
\text{d}t\text{,} \\
\overline{\Delta n_{j}} &=&\frac{1}{t_{f}}\int_{0}^{t_{f}}\Delta n_{j}\left(
t\right) \text{d}t\text{,}
\end{eqnarray}%
where $t_{f}$ represents the cutoff quench time. Average $\overline{C}_{r}$
and $\overline{\Delta n_{j}}$ as functions of parameter $V$ for different $r$
values are plotted in Fig. \ref{fig5}. It can be shown that the peaks appear
in both $\overline{C}_{r}$ and $\overline{\Delta n_{3}}$ when resonant filed
is taken. Comparing to the free fermion case, the increase of such two
quantities suggests that the system experiences a transition from Mott
insulating to conducting phase. This scheme presents an alternative approach
to control the conductivity of the system from the dynamical prospective.

\section{Summary}

\label{summary} In summary, we investigate how does the interplay between
the external field and on-site interaction $U$ alter the conductivity of the
system. When the resonant field is applied, the two-particle case shows that
there can exist an energy shell of $U$, which allows the movement of BP with
the hopping energy scale of $J$ rather than $J^{2}/U$. This phenomenon can
be extended to the dilute system. However, for the case of a single
particle, the movement of the particle is limited due to the existence of
the staggered field, which is reflected by the fact that the bandwidth of
the single-particle spectrum is suppressed from $J$ to $J^{2}/U$ order.
These facts can be examined by the BO. Specifically, when two particles are
bound together, they show large-scale BO in coordinate space. On the
contrary, when the initial two particles are far away, the dc electric field
suppresses the two particles at the initial position. This unique
two-particle property paves the way to understand the many-body quench
dynamics. For the system at half filling, the cooperation of resonant
staggered field and on-site interaction forces the initial antiferromagnetic
Mott insulating state to become a steady conducting state. Our finding
provides a new perspective to understand the influence of external fields on
strongly correlated systems and suggests a new dynamical mechanism to
control the conductivity.

\acknowledgments We acknowledge the support of the National Natural Science
Foundation of China (Grants No. 11975166, and No. 11874225).


\end{document}